\documentclass[conference]{IEEEtran}
\IEEEoverridecommandlockouts
\usepackage{cite}
\usepackage{amsmath,amssymb,amsfonts}
\usepackage{algorithmic}
\usepackage{graphicx}
\usepackage{textcomp}
\usepackage{xcolor}
\def\BibTeX{{\rm B\kern-.05em{\sc i\kern-.025em b}\kern-.08em
    T\kern-.1667em\lower.7ex\hbox{E}\kern-.125emX}}
\begin{document}

\title{Event Detection in Twitter: A Content- and Time-Based Analysis*\\
}

\author{\IEEEauthorblockN{1\textsuperscript{st} Izzat Alsmadi}
\IEEEauthorblockA{\textit{Dept. of computing and cyber security} \\
\textit{Texas A\&M, San Antonio}\\
San Antonio, Texas \\
ialsmadi@tamusa.edu}
\and
\IEEEauthorblockN{2\textsuperscript{nd} Michael J. O’Brien}
\IEEEauthorblockA{\textit{Office of the Provost} \\
\textit{Texas A\&M, San Antonio}\\
San Antonio, Texas \\
mike.obrien@tamusa.edu}
}

\maketitle

\begin{abstract}
The detection of events from online social networks is a recent, evolving field that attracts researchers from across a spectrum of disciplines and domains. Here we report a time-series analysis for predicting events. In particular, we evaluated the frequency distribution of top n-grams of terms over time, focusing on two indicators: high-frequency n-grams over both short and long periods of time. Both indicators can refer to certain aspects of events as they evolve. To evaluate the model’s accuracy in detecting events, we built and used a Twitter dataset of the most-popular hashtags that surrounded the well-documented protests that occurred at the University of Missouri (Mizzou) in late 2015. 
\end{abstract}


\section{Introduction}
\label{intro}
Users create an enormous amount of content through online social networks (OSNs), and tracking and extracting events from such content solely through human labor is impossible. Time is often critical with respect to certain types of events (e.g., security or safety incidents), so there is a need for methods that can be used to track and create alerts for events that users talk about or report through OSNs. Content in OSNs is driven by individual activities or events/news relevant to many users, and the extraction of knowledge regarding contexts and events rests on being able to aggregate information beyond a single activity content or user-account activity. 
Events themselves can be placed in different categories, such as type of event or whether it was identified using supervised or unsupervised techniques \cite{atefeh2015survey}. In the former, dictionaries or datasets of recent significant events can be used to help classifiers make their choice of labels from a pool of alternatives \cite{Sceller}. If event detection is supervised or based on a predefined list of events, keyword-based filtering can be used (\cite{Sakaki}, \cite{Kitagawa} ). Similarly, Twitter streams can be collected as either topic streams, which can be used as the tweets label events, or random streams, which have no specific topics \cite{Kumar}. Event-detection techniques in Twitter face similar challenges related to the limitation of text size, spam, and other types of irrelevant contents, along with the use of abbreviations, slang language, and OSN-specific terms \cite{Weiler}.
Our focus here is on social disruption, especially ones that are sudden. How can normal events turn into disruptive events? Some national or international events that turn into wide-spread disruption can occur on other occasions without triggering similar large publicity. Why? In data-analytic research, assumptions are made that disruptive events can be categorized through temporal and textual features, (\cite{Alsaedi}, \cite{Miro-Llinares}). In reality, however, social and political factors can also contribute to such disruptive events. Those events are not pre-planned, not expected, and mostly unwelcome [9]. They can be triggered by natural disasters and by political or social events at regional, national, and international levels. 

\section{Related Work}
\label{Related Work}
Research in event detection in OSNs can be divided into several categories, two of which are discussed here because of their relevance to our analysis. One is a word or word-pair analysis: (\cite{Kleinberg}, \cite{SayyadiM}, \cite{Weng}, \cite{Ozdikis}, \cite{Zhou}, \cite{Hossny}), where an event can be detected by following certain keywords that show sudden bursts in frequency \cite{Kleinberg}. Trends on Twitter, for example, follow bursts of features or keywords that frequently occur together \cite{Mathioudakis}. Hossny and Mitchell \cite{Hossny} evaluated the time series of the occurrence of word pairs and their association with relevant events. We adopted a similar approach but extended it beyond word pairs to an n-number of words.
The second category of event-detection methods focuses on topics-based detection (\cite{Kontostathis}, \cite{Naaman}, \cite{Long}). On Twitter, topics can be categorized by a hashtag or a mention. Whereas a hashtag is any keyword preceded by a hash sign, a mention is a tweet that contains another user’s @username anywhere in its body. The analysis of mentions and hashtags helps find users who have similar interests. This is very important for many data analytics-based applications besides event detection, such as studying types and natures of interactions and groups among users in OSNs. Users can “favorite” and retweet the posts of other users as well as engage in conversations using @mentions, replies, and hashtags. Events are ranked based on (1) the average “burstiness” of topics  \cite{Alvanaki}, (2) the average usage of emergent topic terms \cite{Cataldi}, or (3) latent Dirichlet allocation \cite{Shi}. Events can be detected based on analyzing user relationships \cite{Du}.
Ideally, if we’re using Twitter, we need an event-detection model based on tweet segments or related terms that occur more often (e.g., more than a threshold level). For disruptive-events detection, the relevant segments should contain more than one term. Further, they should exist in many tweets from many accounts (e.g., include thresholds on the minimum number of tweets and number of accounts). They should also exist on more than certain consecutive time units (e.g., more than a threshold of days). They also need to show up in the top m segments for more than n consecutive time units. Search engines can be used to confirm where segments are popular (e.g., above a certain threshold of search results). A “stickiness” function is defined for optimal segmentation of tweets based on three factors: length normalization, the segment’s presence in Wikipedia, and the segment’s “phraseness,” or the probability of being a phrase based on global and local contexts \cite{Li}. Relevant research employs similar models for detecting “hot” topics (\cite{Zheng}, \cite{Ishikawa}, \cite{Yang}, \cite{Ai}, \cite{Liqing}. Some studies indicate that hot-topics detection can be used to identify rumors and other types of false information in OSNs. Bursty event detection is used to detect hot topics and events that arise quickly, especially in OSNs.

\section{The Event-Detection Model Based on Tweet Segmentation}
\label{The Event-Detection Model Based on Tweet Segmentation}
Our event-detection model is based on the following methods and constraints:
\begin{itemize}
	\item 	Using Tweet segmentation to split a tweet into a sequence of consecutive n-grams, each of which is called a segment. A segment can be a named entity (e.g., a movie title, such as “Finding Nemo”) or a semantically meaningful information unit \cite{Li}.
	\item	Hashtags can be a good starting point, but a hashtag is only one word without spaces. The relevant segments should contain more than one term. On the other hand, we noticed that in a dataset of related tweets, hashtags will appear as popular words or n-grams. As a result, we evaluated the use of hashtag segmentation for event detection (\cite{Berardi}, \cite{Park}, \cite{Celebi}).
	\item	Segments should exist in many tweets from many accounts (thresholds on the number of tweets and number of accounts).
	\item	Segments should also exist on more than a few consecutive days; they need to show up in the top-10 segments for more than n consecutive days.
	\item	Segments should be popular or trending in search engines or on social networks within days of interest.
\end{itemize}

\begin{figure}
	\includegraphics{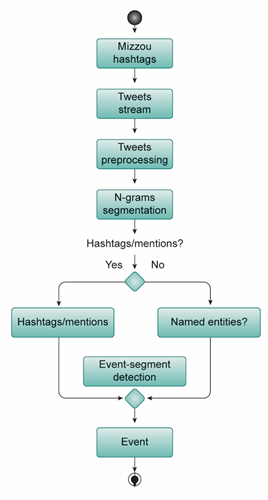}
	\caption{Twitter event-detection architecture}
	\label{fig:1}       
\end{figure}

Fig. \ref{fig:1} shows our proposed overall architecture for Twitter-based event detection based on initial hashtags.


\section{Experiments and Analysis}
\label{Experiments and Analysis}

As the subject of this research is not a current event, and no public dataset from Twitter can be used, we collected a set of related tweets based on Twitter public API
(https://developer.twitter.com/en/docs/twitter-api). In order to ensure that collected tweets were relevant to the Mizzou protests, we used a selection of popular hashtags related to the event as an input to the tweet-collection process. We selected a period of approximately a year around the protests and their immediate aftermath (November 1, 2015, to December 31, 2016).

\subsection{Mizzou protests popular hashtags}
\label{Mizzou protests popular hashtags}
We defined the following four hashtags as being the most relevant to the event: \#PrayForMizzou, \#ConcernedStudent1950, \#BlackLivesMatter, and \#Mizzou. Table \ref{tab:1} shows basic statistics on the tweets collected as a result of those hashtags. The table also shows the average positive and negative polarity of the sentiment in the tweets for each hashtag found using TextBlob sentiment library (vol. 15.3). 


\begin{table}[htbp]
	\caption{Top hashtags surrounding the Mizzou 2015 protests}
	\label{tab:1}       
	\resizebox{\columnwidth}{!}{\begin{tabular}{|c|c|c|c|}
		\hline\noalign{\smallskip}
		Hashtag &	NO. of Tweets &	Avg Pos. Sentiments &	Avg Neg. entiments \\
		\noalign{\smallskip}\hline\noalign{\smallskip}
		
		\#PrayForMizzou	 & 10,000 &	0.537517 &	0.462483 \\
		\#MizzouHungerStrike &	6,514 &	0.632995 &	0.367005 \\
		\#Mizzou &	71,630 &	0.550789 &	0.449211 \\
		\#BlackLivesMatter &	71,421 &	0.550862 &	0.449138 \\

		\noalign{\smallskip}\hline
	\end{tabular}}
\end{table}

\begin{figure}
	\resizebox{\linewidth}{!}{\includegraphics[width=0.5\linewidth]{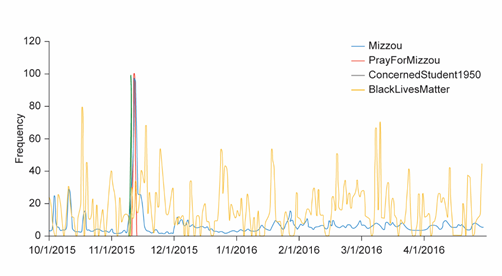}}
	\caption{Timeline for four hashtags related to the Mizzou protests}
	\label{fig:2}       
\end{figure}

\begin{figure}
	\resizebox{\linewidth}{!}{\includegraphics{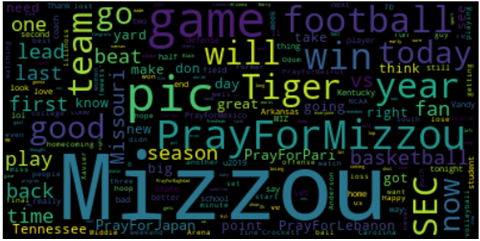}}
	\caption{Mizzou-related tweets word cloud}
	\label{fig:3}       
\end{figure}

Fig. \ref{fig:2} shows the hashtags’ timelines. The timelines for two hashtags—\#PrayForMizzou and \#ConcernedStudent1950—are restricted to November 2015, with another—Mizzou—peaking during that month but lasting throughout the timeline. \#BlackLivesMatter fluctuates randomly throughout the timeline. The timelines of the hashtags can show indications of how the focus or the subject of the protest evolves. Fig. \ref{fig:3} shows a word cloud for the top single terms in the dataset that contain tweets from all hashtags. The cloud shows major terms from the protests as well as from sports-related activities.


\subsection*{N-grams-based tweets segmentation}
\label{N-grams-based tweets segmentation}

In order to extract events from a large corpus of text, we studied patterns of statements and phrases. Our goal was to segment tweets into n-grams of words. We limited our experiments to n-grams of 2-5 words. Table \ref{tab:2} shows the top-20 bi-gram terms. The majority of those share the term “Mizzou” in common, but little in terms of relevant events can be extracted from those bi-grams. Table \ref{tab:3} shows the top-20 tri-grams. Here we can start observing more possibilities for extracting significant events. Tables \ref{tab:4} and \ref{tab:5} shows grams for four and five terms, respectively. We can see that with some necessary cleaning, those grams will be better candidates to detect significant events in the tweets dataset. Typical text analysis pre-processing includes removing stop words, which are commonly used words such as articles, pronouns, and prepositions. Each language typically includes a large list of stop words that can be used. In our experiment, we did not employ stop-word removal in pre-processing stages, as our goal was to search for tweet segments that could serve as phrases or statements in the event-detection process. 


\begin{table}
	\caption{Top bi-grams surrounding the Mizzou 2015 protests}
	\label{tab:2}       
	\resizebox{\linewidth}{!}{\begin{tabular}{llll}
		\hline\noalign{\smallskip}
		Bi-gram &	Count &	Bi-gram &	Count \\
		\noalign{\smallskip}\hline\noalign{\smallskip}

		pic twitter &	14031 &	i m &	4737 \\
		twitter com &	14031 &	it s &	4635 \\
		at mizzou &	9878 &	on the &	4497 \\
		in the &	8050 &	mizzou https &	4222 \\
		mizzou s &	6577 &	for the &	3877 \\
		the mizzou &	6455 &	to the &	3738 \\
		to mizzou &	6228 &	mizzou mizzou &	3711 \\
		for mizzou &	6206 &	mizzou football &	3566 \\
		of the &	5770 &	don t &	3360 \\
		mizzou is &	5644 &	at the &	3346 \\

		\noalign{\smallskip}\hline
	\end{tabular}}
\end{table}

\begin{table}
	\caption{Top tri-grams surrounding the Mizzou 2015 protests.}
	\label{tab:3}       
	\resizebox{\linewidth}{!}{\begin{tabular}{llll}
		\hline\noalign{\smallskip}
		Tri-gram &	Count &	Tri-gram &	Count \\
		\noalign{\smallskip}\hline\noalign{\smallskip}

		pic twitter com &	14031 &	let s go &	597 \\
		mizzou pic twitter&	2220 &	to be a	&568\\
		at mizzou arena	&1234&	miz pic twitter&	537 \\
		the mizzou game&	1146&	go to mizzou	&530 \\
		i don t&	1026&	i can t	&508 \\
		a mizzou fan&	931&	it s a	&505 \\
		in the sec&	753	&prayforbeirut prayformexico prayformizzou &	500 \\
		at the mizzou&	712&	prayforparis prayforjapan prayforbeirut	&497 \\
		to the mizzou&	666&	is going to	&485 \\
		mizzou https mizzou&	661&	can t wait	&483 \\

		\noalign{\smallskip}\hline
	\end{tabular}}
\end{table}

\begin{table}
	\caption{Top four-grams surrounding the Mizzou 2015 protests.}
	\label{tab:4}       
	\resizebox{\linewidth}{!}{\begin{tabular}{llll}
		\hline\noalign{\smallskip}
		Four-gram &	Count &	Four-gram &	Count \\
		\noalign{\smallskip}\hline\noalign{\smallskip}

		mizzou pic twitter com &	2220 &	phenom rb arrested hours &	367 \\
		miz pic twitter com	&537	&rb arrested hours after	&367\\
		prayforjapan prayforbeirut prayformexico prayformizzou&	442&	arrested hours after breaking&	355 \\
		prayforparis prayforjapan prayforbeirut prayformexico	&399&	breaking school records mug	&348 \\
		freshman phenom rb arrested&	388&	school records mug shot	&327 \\
		prayformizzou pic twitter com	&387&	mizzou tweets mizzou tweets&	324 \\
		hours after breaking school	&372&	can t wait to	&300 \\
		after breaking school records&	372	&let s go mizzou	&294 \\
		mizzou football freshman phenom	&370&	she took exams for	&294 \\
		football freshman phenom rb	&370	&took exams for mizzou&	285 \\

		\noalign{\smallskip}\hline
	\end{tabular}}
\end{table}

\begin{table}
	\caption{Top five-grams surrounding the Mizzou 2015 protests.}
	\label{tab:5}       
	\resizebox{\linewidth}{!}{\begin{tabular}{llll}
		\hline\noalign{\smallskip}
		Five-gram &	Count &	Five-gram &	Count \\
		\noalign{\smallskip}\hline\noalign{\smallskip}

		prayforparis prayforjapan prayforbeirut prayformexico prayformizzou	&390&	she took exams for mizzou	285 \\
		hours after breaking school records	&372&	took exams for mizzou athletes	&285 \\
		mizzou football freshman phenom rb	&370&	missouri tigers vs az wildcats&	282 \\
		football freshman phenom rb arrested	&367&	school records mug shot http&	273 \\
		freshman phenom rb arrested hours	&367&	tutor says she took exams	&267 \\
		phenom rb arrested hours after	&367&	says she took exams for	&267 \\
		rb arrested hours after breaking	&355&	dec 10 10 missouri tigers&	264 \\
		arrested hours after breaking school	7351&	10 10 missouri tigers vs	&264 \\
		after breaking school records mug	&348&	10 missouri tigers vs az	&264 \\
		breaking school records mug shot&	327&	prayforjapan prayforbeirut prayformexico prayformizzou prayforlebanon	&243 \\

		\noalign{\smallskip}\hline
	\end{tabular}}
\end{table}

\begin{figure}
	\resizebox{\linewidth}{!}{\includegraphics{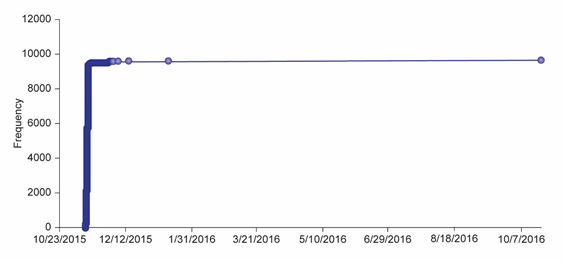}}
	\caption{\#PrayForMizzou time series. The dots represent a few mentions after December 2015. }
	\label{fig:4}       
\end{figure}

\begin{figure}
	\resizebox{\linewidth}{!}{\includegraphics{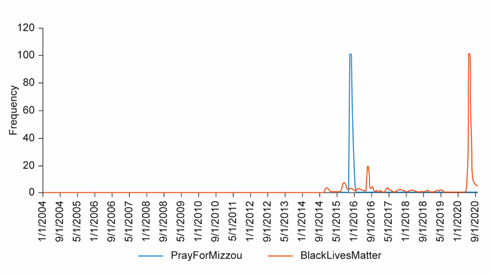}}
	\caption{\#PrayForMizzou versus \#BlackLivesMatter time series}
	\label{fig:5}       
\end{figure}

Time series for hashtags and popular terms can help us visualize their evolution in terms of frequency and dates/times. The large increase in some terms within the short term is what we will use in events’ prediction. Fig. 4 shows a sample time series for the term “PrayForMizzou,” including but not limited to the hashtag \#PrayForMizzou. Fig. \ref{fig:4} shows a rise in the term of almost 10,000 times within roughly a one-month period (November 2015). Fig. \ref{fig:5} shows correlated results between Twitter and Google Trends. We compared the time series of \#PrayForMizzou vs. \#BlackLivesMatter. Fig. 5 shows that unlike BlackLivesMatter, the term PrayForMizzou is associated with only a single event (Mizzou protests). As hashtags are specially formed terms—a single word formed from several words—they can be uniquely associated with some events.


\begin{figure}
	\resizebox{\linewidth}{!}{\includegraphics{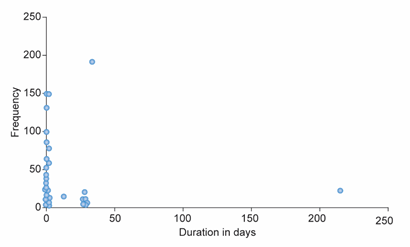}}
	\caption{Duration versus count for three-gram hashtags}
	\label{fig:6}       
\end{figure}

We proposed three n-gram-based metrics that we wanted to evaluate in terms of their ability to detect events:

\begin{itemize}
	\item Total count of top n-grams: Our observations as well as our literature review indicated that significant events would be mentioned by many users. Our assumption was that we could find some phrases that were repeated in high volumes and that could be associated with certain events.
	\item	Duration: Our previous analysis of hashtags and Google Trends showed that studying the frequencies or trends of certain phrases can be associated with certain events. “Duration metric” refers to the difference in time between the first and the last n-gram in the dataset.
	\item	Density: This is our key indicator to show the burst of the event by dividing the total number of identical n-grams by the duration.
	Our detection model is built around the analysis of a triangle of Twitter accounts, tweets n-grams’ frequencies, and tweets n-grams’ duration. 
	Fig. 6 shows both frequencies and duration for the top 138 three-grams. Based on Fig. 6 we identified two categories of candidate n-grams:
	\item	A very high number of n-grams repeated over a short duration. Those would be the data points that extend up the y-axis in Fig. 6. 
	\item	A lower number of n-grams repeated over a longer duration. Those would be the ones that occur around day 35 in Fig. \ref{fig:6} Those are second-level candidates for detecting events, as they need to be evaluated further (e.g., using metrics in addition to duration and frequency).

\end{itemize}

\section{Conclusion}
The power and influence of Online Social Networks (OSNs) on our daily lives are beyond question. People nowadays use those networks for social gatherings and activities, movements, protests, and the like. Our effort was focused on evaluating a model to detect and predict large-scale events based on OSNs:

\begin{itemize}
	\item We proposed and evaluated metrics that can be automatically collected based on tracking certain inputs such as hashtags and Google Trends. 
	\item	We evaluated our model based on the University of Missouri protests of 2015 and the aftermath. 
	\item	We collected a large dataset of tweets around the events based on some of the more popular hashtags used during that period. 
	\item	We focused on top n-grams to generate candidate events. 
	Our results showed that n-grams of words are better than hashtags for predicting the occurrence of events. There are several reasons for this. First, top n-grams of words show popular word sequences, whereas hashtags represent single words or phrases. Second, popular hashtags can be included in popular n-grams and n-grams can show us combinations of popular hashtags related to each other. For example, one of the popular three-grams is (prayformexico prayformizzou prayforlebanon), which shows three popular, related hashtags that reflect events in different parts of the world.
\end{itemize}

We proposed simple but effective metrics to calculate automatically what can help predict the occur of events: n-gram frequency versus time. Based on the relation between frequency and time, we identified two types of interesting types: (1) high n-gram frequencies within a short time—a few hours or a few days—and (2) high n-gram frequencies over a longer time—a few months, for example. We think that both types can be good event predictors. And we need good predictors because by the time we begin to sense that an event is upon us, it might be too late to escape the aftermath. Take the case of the University of Missouri. It took a little over a year for the fuse lit by the August 2014 killing of Michael Brown in Ferguson, Missouri, to lead to the first disruption in September 2015—a disruption that immediately led to more—and more volatile—disruptions \cite{Pearson}.
The toll exacted by disruptive events can be significant. There is no doubt of the psychological and cognitive fatigue caused by trolling, not to mention the possibility of financial disaster. For example, in the two years following the Mizzou protests, the university saw a 35\% drop in freshman enrollment and an overall drop of 14\%, causing the campus to cut 12\%—about \$55 million—from the academic and administrative operations budget \cite{Keller}. This, in turn, led to significant layoffs of faculty and staff. Even today, the campus is nowhere back to what it was before the protests, neither financially, socially, nor politically. The take-home message is clear: We need to continue to develop solid methods for predicting disruptive events.

\subsection{Note}
One of us, MOB, was dean of the College of Arts and Science at Mizzou during the unrest and saw firsthand the protest and aftermath unfold.

\section*{Acknowledgment}
We thank Gloria O'Brien for her editorial assistance.

\vspace{12pt}

\end{document}